\documentclass{aa}
\usepackage{graphicx,natbib}
\bibpunct{(}{)}{;}{a}{}{,}
\begin{document}
   \title{On the use of $^{12}$CO/$^{13}$CO as a test of common-envelope evolution}

   \subtitle{}

   \author{V. S. Dhillon
          \inst{1}
          \and
          S. P. Littlefair
          \inst{1}\fnmsep\thanks{Present address: Astrophysics Group, 
          School of Physics, University of Exeter, Stocker Road,
          Exeter, EX4 4QL, UK ({\tt sl@astro.ex.ac.uk}).}
          \and
	  T. R. Marsh
          \inst{2}
          \and
	  M. J. Sarna
          \inst{3}
          \and
	  E. H. Boakes
          \inst{1}
          }

   \offprints{V. S. Dhillon}

   \institute{Department of Physics and Astronomy, University of Sheffield, 
              Sheffield, S3 7RH, UK\\
              \email{vik.dhillon@sheffield.ac.uk}
         \and
             Department of Physics and Astronomy, University of Southampton, 
             Highfield, Southampton, SO17 1BJ, UK\\
             \email{trm@astro.soton.ac.uk}
         \and
             N. Copernicus Astronomical Center, Polish Academy of Sciences, 
             ul. Bartycka 18, 00-716 Warsaw, Poland\\
             \email{sarna@camk.edu.pl}
             }

   \date{Received September 7, 2001; accepted July 16, 2002}

   \abstract{ We present K-band echelle spectra of the cataclysmic
variable SS~Cyg and the pre-cataclysmic variable V471 Tau in order to
measure the strengths of the $^{12}$CO and $^{13}$CO bands at 2.3525
and 2.3448 $\mu$m, respectively, and so perform the observational test
of the common-envelope model of close binary star evolution proposed
by \citet{sarna95}. Although we find evidence of an absorption feature
coincident with the expected wavelength of $^{13}$CO in both objects,
we attribute it instead to a cluster of neutral atomic absorption
features (primarily due to Ti\,{\small I}) possibly arising from
star-spots on the surfaces of the rapidly rotating secondary stars in
these systems, thereby rendering the test inconclusive. We present a
modified observational test of common-envelope evolution, based on the
observation of the $^{13}$CO bands at 2.3739 and 2.4037 $\mu$m, which
is insensitive to spectral contamination by star-spots.
\keywords{binaries: spectroscopic -- stars: individual: SS Cyg --
stars: individual: V471 Tau -- novae, cataclysmic variables --
infrared: stars -- nuclear reactions, nucleosynthesis, abundances } }

   \maketitle
%
%________________________________________________________________

\section{Introduction}

Well over half of all stars are believed to be binary or multiple
systems, with about half of these, in turn, consisting of close binary
systems where the two component stars are unable to complete their
normal evolution without being influenced by the presence of the other
(see \citet{duquennoy91} and references therein).  The orbital
separations of close binary systems containing at least one compact
object -- such as the cataclysmic variables (CVs) and low-mass X-ray
binaries (LMXBs) -- are significantly smaller than the radii of the
stars which were the progenitors of the compact objects in these
systems. This means that significant orbital shrinkage must have
occurred, probably in a process known as common-envelope (CE)
evolution. According to the CE model of close binary star evolution
(in this case, as applied to CVs; \citealt{paczynski76}), the more
massive (primary) star fills its Roche lobe when it reaches its giant
or asymptotic giant branch phase, while its lower mass (secondary)
companion remains on the main sequence. Under these conditions, mass
transfer to the secondary is dynamically unstable and occurs at such a
high rate that the transferred material cannot be accreted by the
secondary and so forms a CE, in which the binary is immersed. Through
the action of drag forces, the main-sequence star spirals
towards the core of the giant, generating luminosity which drives off
the CE.  What remains is often called a post-common envelope binary
(PCEB), with typical orbital periods of a few days.  These systems are
thought to become CVs or LMXBs when magnetic braking or gravitational
radiation extracts sufficient orbital angular momentum for the
main-sequence secondary star to fill its Roche lobe
(\citealt{spruit83}; \citealt{rappaport83}). The theory of CE
evolution is therefore of fundamental importance in astrophysics, and
is probably a key step in the production of some of the most exotic
inhabitants of our Galaxy, including the binary radio pulsars,
black-hole X-ray binaries and Type Ia supernovae. For a recent review
of CE evolution, see \citet{iben93}.

Although there is general agreement that most close binary systems
have evolved through a CE phase, there is little direct evidence to
support the CE model.  The best evidence to date for the reality of CE
evolution comes from the observation of planetary nebulae with close
binary nuclei (\citealt{iben93}; \citealt{livio96}; \citealt{bond92}).
There is no direct evidence, however, that whole classes of important
objects such as LMXBs, CVs and their immediate precursors, the
so-called pre-CVs (e.g. \citealt{catalan94}), have in fact evolved
through a CE phase.  As a result, \citet{sarna95} proposed a direct
observational test of CE evolution. The idea is based on the fact that
the ratio of $^{12}$C/$^{13}$C decreases from a value of 84 in
main-sequence stars like the Sun \citep{harris87} to a value of about
17 in giants \citep{harris88}, due to the different nuclear burning
and mixing processes which occur in these stars. During the CE phase,
the main-sequence secondary effectively exists within the atmosphere
of the giant primary and will accrete material from it, thereby
altering the $^{12}$C/$^{13}$C ratio from solar-like values towards
giant-like values. By measuring the $^{12}$C/$^{13}$C ratio it is
therefore possible to determine whether a binary has passed through a
CE stage.  This test has already been performed by \citet{dhillon95a},
who made a tentative detection of $^{13}$CO in the K-band spectrum of
the pre-CV V471~Tau. To confirm this detection, we observed V471 Tau
again, along with the CV SS Cyg. The results of these new, much higher
quality observations are presented in this paper, together with a
discussion of the recent results of \citealt{catalan01} (who
independently performed similar observations to the ones we describe
here).

%__________________________________________________________________
%

\section{Observations \& Data Reduction}

On the night of 1995 October 22 we obtained 2.3338--2.3662~$\mu$m
echelle  (15 km\,s$^{-1}$ resolution) spectra of the detached
binary/pre-CV V471 Tau,  the semi-detached binary/CV  SS Cyg  and the
field stars Gl105A (a K3 dwarf) and BS86 (a K3 giant) with CGS4
\citep{wright94} on the 3.8 m United Kingdom Infrared Telescope (UKIRT)
on Mauna Kea, Hawaii. This was followed by additional UKIRT+CGS4
observations on the night of 2000  June 19, when we obtained
2.3310--2.3481~$\mu$m echelle spectra of the field stars Gl406 (an M6
dwarf) and GJ1190 (a K5 dwarf) at a resolution of 8 km\,s$^{-1}$.
Regular observations of nearby standards with featureless spectra were
obtained on both nights to correct for the effects of telluric
absorption and to provide flux calibration. The data were reduced
following the procedures described  by \citet{dhillon95a}.

%__________________________________________________________________
%

\section{Results}

\begin{figure*}
\centering
\includegraphics[width=12cm]{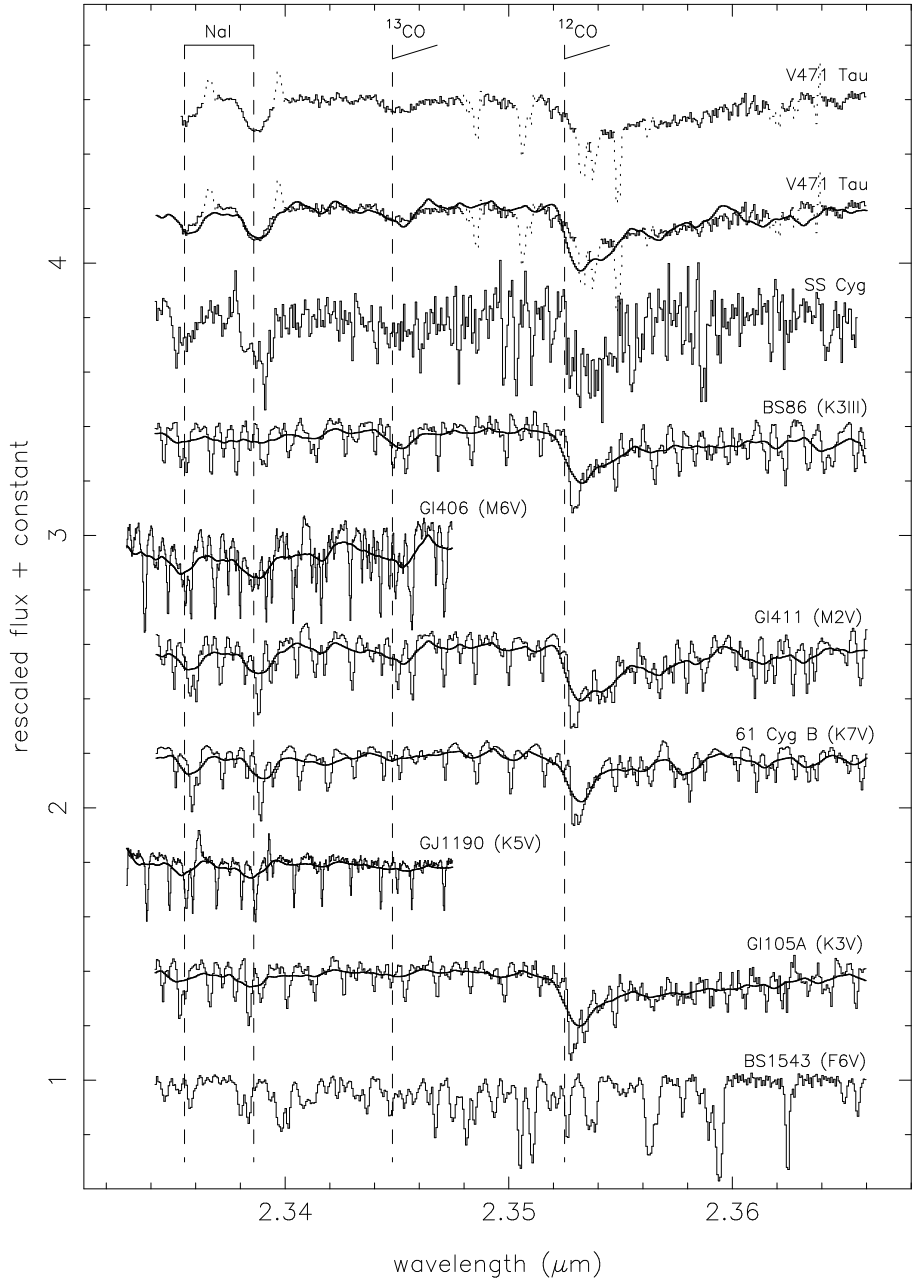}
\caption{Averaged K-band echelle spectra of (from top to bottom) the
pre-CV V471 Tau (top two spectra), the CV SS Cyg, a K3 giant (BS86),
and dwarf stars of spectral type M6--K3. The spectra of Gl411 and 61
Cyg are taken from the atlas of \protect\citet{wallace96}, and have
been binned to exhibit the same resolution as the other spectra.
All spectra have been normalized by dividing by a spline fit to their
continua. Despite the number of lines in each spectrum, identifying
the continuum was not difficult because it is essentially flat due to
the short wavelength range covered. To account for the contribution of
non-secondary star light in the binaries, the spectra have also been
scaled to ensure that the $^{12}$CO bands are of an equivalent depth
(thereby enabling a direct comparison of the $^{12}$CO/$^{13}$CO
ratios in each object). All spectra have been offset on the ordinate
by adding a constant to each spectrum.  The bottom spectrum is that of
an F6 dwarf (BS1543), normalized by dividing by a spline fit to the
continuum regions, which indicates the location telluric absorption
features. The thick, solid curves superimposed on the spectra of the
dwarf and giant stars show these spectra rotationally broadened by 90
km\,s$^{-1}$. The thick, solid curve superimposed on the lower
spectrum of V471~Tau is the rotationally broadened spectrum of the M2
dwarf Gl411. The dotted lines in the spectrum of V471 Tau denote bad
pixels and regions where the telluric correction was poor.}
\label{fig:spectra}
\end{figure*}

The observed spectra are presented in Fig.~\ref{fig:spectra}. The
spectral features of interest are the first-overtone,
vibrational-rotational molecular bands of $^{12}$CO (4--2) at 2.3525
$\mu$m and $^{13}$CO (2--0) at 2.3448 $\mu$m, where the wavelengths
refer to the position of the band-head. Given that the ratio of
$^{12}$C/$^{13}$C is approximately 84 in solar-like stars and 17 in
giants, one would expect $^{13}$CO to be significantly stronger in the
spectrum of a K3III star than in a K3V star. This is precisely what is
observed in Fig.~\ref{fig:spectra} -- the $^{12}$CO band is prominent
in both Gl105A and BS86, but the $^{13}$CO band does not appear in the
former whereas it is clearly visible in the latter (see also
Table~\ref{tab:lines}).  The spectrum of the detached binary V471 Tau
in Fig.~\ref{fig:spectra} also shows prominent $^{12}$CO. This is to
be expected, given that this 12.5 hr-period pre-CV has a secondary
star with an estimated spectral type of K2V
\citep{young72}\footnote{This spectral type estimate is far from
accurate, however, being based solely on the U-B and B-V colours}. The
spectrum of V471 Tau also shows a prominent absorption feature
coincident with the expected wavelength of $^{13}$CO, with a strength
relative to $^{12}$CO similar to that observed in BS86 (see
Table~\ref{tab:lines}). This is in agreement with the detection of
such a feature by \citet{dhillon95a} and \citet{catalan01}. The
absorption feature is also observed in the spectrum of SS Cyg in
Fig.~\ref{fig:spectra}; the (albeit noisier) spectrum of this 6.6
hr-period CV with a K4V secondary star \citep{ritter98} shows enhanced
absorption around 2.3448 $\mu$m, once again in agreement with the
somewhat uncertain detection by \citet{catalan01}.

\begin{table*}
\centering
\caption{ Flux deficits of the molecular bands in V471 Tau, SS Cyg and
two representative spectral-type templates. The flux deficits were
measured by integrating the flux in the continuum-subtracted spectra
between 2.3445--2.3465 $\mu$m for $^{13}$CO and between 2.3515--2.3600
$\mu$m for $^{12}$CO (with an appropriate correction for the Doppler
shift in V471 Tau and SS Cyg).  Note that the spectra suffer from slit
losses and hence the individual flux deficits are only lower limits;
the flux deficit ratios given in the right-hand column, however, are
secure. Equivalent width measurements are not presented due to
contaminating emission from the white dwarf and the effects of
irradiation in V471 Tau and SS Cyg.}
\begin{tabular}{lccc}
\label{tab:lines}
& & \\
\hline
& & \\
Object   & $^{13}$CO & $^{12}$CO & $^{13}$CO/$^{12}$CO \\ 
         & {\small ($\times$\,10$^{-13}$\,erg\,s$^{-1}$\,cm$^{-2}$)} 
         & {\small ($\times$\,10$^{-13}$\,erg\,s$^{-1}$\,cm$^{-2}$)} & \\
         &                                                           & \\
V471 Tau & \ 0.122 $\pm$ 0.008$^*$ & 1.331 $\pm$ 0.017 & \ 0.092 $\pm$ 0.006$^*$ \\
SS Cyg   & \ 0.015 $\pm$ 0.002$^*$ & 0.123 $\pm$ 0.006 & \ 0.122 $\pm$ 0.017$^*$ \\
Gl105A (K3V) & 0.019 $\pm$ 0.003 & 1.290 $\pm$ 0.006 & 0.015 $\pm$ 0.002 \\
BS86 (K3III) & 0.341 $\pm$ 0.004 & 3.520 $\pm$ 0.009 & 0.097 $\pm$ 0.001 \\
& & \\
\multicolumn{4}{l}{$^*$The identification of the absorption band with 
$^{13}$CO in these systems is not secure -- see text for details.} \\
& & \\
\hline
& & \\
\end{tabular}
\end{table*}

One interpretation of the above result is that we have detected
enhanced $^{13}$CO, thereby confirming that V471 Tau and SS Cyg have
evolved through a CE phase. This interpretation would almost certainly
be incorrect, however, as the spectra of both objects show evidence of
contamination by star-spots. This can be understood by inspecting the
Na\,{\small I} features at 2.3355 $\mu$m and 2.3386 $\mu$m in
Fig.~\ref{fig:spectra}. The Na\,{\small I} doublet in both V471 Tau
and SS Cyg is stronger than one would expect given the spectral type
of the secondary stars in these objects (compare the strength of
Na\,{\small I} in V471~Tau and SS~Cyg with its strength in the K3V
star in Fig.~\ref{fig:spectra}). It is straightforward to explain this
excess Na\,{\small I} absorption as being due to star-spot
contamination, as the cooler star-spots would contribute to spectral
features seen in stars of a later spectral type (and hence strengthen
Na\,{\small I} in the spectra of V471 Tau and SS Cyg; compare the
strength of Na\,{\small I} in the K3V spectrum with the K7V and M2V
spectra in Fig.~\ref{fig:spectra}). This interpretation is supported
by the Doppler imaging experiments of \citet{ramseyer95}, who find
that the secondary star in V471 Tau exhibits extensive star-spot
coverage ($\sim25$\% of the stellar surface, according to
\citealt{obrien01}). \citet{webb01} also find evidence for $\sim22$\%
star-spot coverage of the secondary star of SS Cyg. They see a similar
effect to the one described in this paper; in their case, TiO
absorption from star-spots is introduced into the optical spectrum of
the K4 secondary star (and TiO absorption is not normally seen in
stars earlier than K7).

It should be pointed out that star-spots are not the only explanation
for the anomalous strength of Na\,{\small I} absorption in our spectra
of V471 Tau and SS Cyg; differences in temperature, gravity or
abundances between the secondary stars and the spectral-type templates
in Fig.~\ref{fig:spectra} could also cause such an effect. Given that
we already know star-spots are present in V471 Tau and SS Cyg,
however, this seems to be the simplest explanation.  If we accept that
the enhancement in Na\,{\small I} line strength is due to
contamination by star-spots, this then implies that the feature at
2.3448 $\mu$m in Fig.~\ref{fig:spectra} may not be due to $^{13}$CO.
This can be seen by examining the rotationally broadened K7V, M2V and
M6V spectra in Fig.~\ref{fig:spectra}, which shows a prominent
absorption feature around 2.3448 $\mu$m. This absorption feature is
dominated by the neutral atomic line of Ti\,{\small I}
\citep{wallace96} and happens to fall at the same position in the
spectrum (2.3448 $\mu$m) as $^{13}$CO. Furthermore, it can be seen
that the Ti\,{\small I} feature increases in strength with later
spectral type. As a result, the absorption feature at 2.3448 $\mu$m in
V471 Tau and SS Cyg most probably arises at least partly from
Ti\,{\small I}. To illustrate this point we have plotted the
rotationally-broadened spectrum of Gl411, an M2 dwarf star, on top of
the spectrum of V471~Tau in Fig.~\ref{fig:spectra}.  The spectrum of
the M2 dwarf not only provides a good match to the strengths of both
Na\,{\small I} and $^{12}$CO in V471 Tau, but also an adequate match
to the absorption feature at 2.3448$\mu$m. Given the uncertainty in
the spectrum of the star-spots, the fraction of the surface they cover
and the temperature of the surrounding photosphere, it is not possible
to reliably disentangle the Ti\,{\small I} absorption in V471 Tau and
SS Cyg from any $^{13}$CO that may be present. In other words, it is
impossible to reliably distinguish between the effects of CE evolution
and star-spot contamination using the spectra presented in
Fig.~\ref{fig:spectra}.

\begin{figure*}
\centering
\includegraphics[width=12cm]{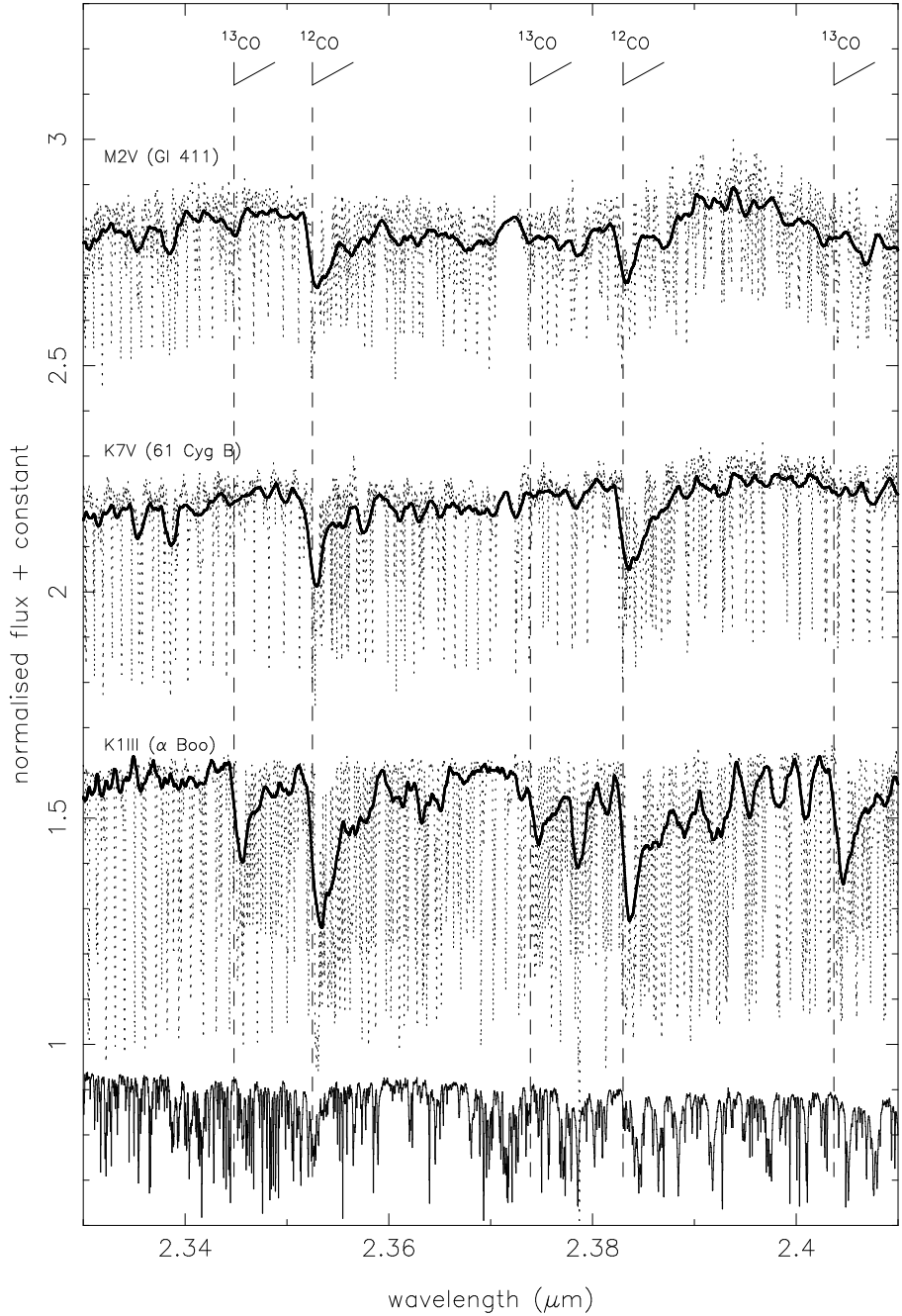}
\caption{The dotted curves show the high resolution K-band spectra of
(from top to bottom) the M2 dwarf Gl 411, the K7 dwarf 61 Cyg B and
the K1 giant $\alpha$ Boo,  taken from the atlas of
\protect\citet{wallace96}. The spectra have been normalised by
dividing by the flux at 2.37 $\mu$m and offset on the ordinate by
adding a  constant to each spectrum. The thick, solid curves
superimposed on the spectra of the dwarf and giant stars show these
spectra  rotationally broadened by 90 km\,s$^{-1}$. The bottom
spectrum is an atmospheric transmission function obtained from the
ratios of two stellar spectra taken at different airmasses -- the
details of this process are given in \protect\citet{wallace96}.}
\label{fig:highres}
\end{figure*}

The test proposed by \citet{sarna95} still remains viable, however.
Figure~\ref{fig:highres} shows high-resolution K-band library spectra
taken from the spectral atlas of \citet{wallace96}. Over-plotted in
thick lines are the same spectra rotationally broadened by 90
km\,s$^{-1}$, corresponding to the projected rotational velocity of
the secondary star in V471~Tau \citep{ramseyer95}. Prominent in the
spectrum of the K1 giant, $\alpha$ Boo, are the absorption bands of
$^{12}$CO at 2.3525 and 2.3830 $\mu$m, as well as the absorption bands
of $^{13}$CO at 2.3448, 2.3739 and 2.4037 $\mu$m. The $^{13}$CO bands
are not present in the spectrum of the K7 dwarf, but the rotationally
broadened M2 dwarf spectrum shows an absorption feature at 2.3448
$\mu$m -- as seen in the spectra presented in
Fig.~\ref{fig:spectra}. However, the rotationally broadened M2 dwarf
spectrum displays no absorption features at wavelengths corresponding
to the $^{13}$CO bands at 2.3739 and 2.4037 $\mu$m. Observations of
V471 Tau and SS Cyg at these wavelengths (particularly the latter,
stronger feature) would therefore be free of star-spot contamination
and would allow a successful test of the CE model.

%__________________________________________________________________
%

\section{A comment on the results of \citet{catalan01}}

Independently of the work present in this paper, \citet{catalan01}
observed V471 Tau, SS Cyg and four other binaries in order to confirm
the results of \citet{dhillon95a}. They found evidence for the
spectral feature at 2.3448 $\mu$m in V471 Tau, SS Cyg and one other
detached binary (Feige 24). They attribute this to a positive
detection of $^{13}$CO; for the reasons discussed above, we believe
this is erroneous and that the feature is most probably due to
star-spots.  The data of \citet{catalan01} cover a wider wavelength
range than the data presented here, including the $^{13}$CO band at
2.3739 $\mu$m.  They have therefore partially performed the new
observational test of CE evolution that we describe in this paper and
they report a detection of the $^{13}$CO band at 2.3739 $\mu$m in the
binary Feige 24. However, their detection is at a very low
signal-to-noise, the feature does not appear to lie at the correct
wavelength (see their Fig.~2), and they do not detect this band in any
of the other binaries observed. Therefore, we believe that
\citet{catalan01} present no strong evidence for the presence of
$^{13}$CO and, unfortunately, their wavelength coverage does not
extend to the much stronger $^{13}$CO feature at 2.4037 $\mu$m that we
propose should be observed in future observational tests.

%__________________________________________________________________
%

\section{Conclusions}

Our K-band echelle spectra of the CV SS Cyg and the pre-CV V471 Tau
show evidence of a spectral feature at 2.3448 $\mu$m. This is coincident
with where we would expect to observe the $^{13}$CO (2--0) molecular
band, which would imply that these objects have passed through a CE
phase during their evolution. However, our spectra
also show enhanced Na\,{\small I} absorption, which can be attributed
to star-spots. These star-spots also contribute to a cluster of strong,
neutral atomic absorption features around 2.3448 $\mu$m, rendering
our test of CE evolution inconclusive. We therefore propose a new test
based on the measurement of the $^{13}$CO bands at 2.3739 and 2.4037 $\mu$m, 
which we show would be uncontaminated by star-spots. We intend to perform
this revised test in the near future.

%__________________________________________________________________
%

\begin{acknowledgements}

We would like to thank the anonymous referee for his/her comments on a
draft of this paper.  UKIRT is operated by the Joint Astronomy Centre
on behalf of the Particle Physics and Astronomy Research Council.  The
authors acknowledge the data reduction and analysis facilities
provided at the University of Sheffield by the Starlink Project which
is run by CCLRC on behalf of PPARC. MJS acknowledges the support of
the State Committee for Scientific Research through grant
2-P03D-005-16.

\end{acknowledgements}

%__________________________________________________________________
%

%__________________________________________________________________
%


\begin{thebibliography}{20}
\expandafter\ifx\csname natexlab\endcsname\relax\def\natexlab#1{#1}\fi

\bibitem[{Bond(1992)}]{bond92}
Bond, H.~E. 1992, in ASP Conf. Ser. 56, 179

\bibitem[{{Catal{\' a}n} {et~al.}(2001){Catal{\' a}n}, {Smalley}, {Exter},
  {Wood}, \& {Sarna}}]{catalan01}
{Catal{\' a}n}, M.~S., {Smalley}, B., {Exter}, K.~M., {Wood}, J.~H., \&
  {Sarna}, M.~J. 2001, in ASP Conf. Ser. 229, 251

\bibitem[{Catal\'{a}n {et~al.}(1994)Catal\'{a}n, Davey, Sarna, Smith, \&
  Wood}]{catalan94}
Catal\'{a}n, M.~S., Davey, S.~C., Sarna, M.~J., Smith, R.~C., \& Wood, J.~H.
  1994, MNRAS, 269, 879

\bibitem[{Dhillon \& Marsh(1995)}]{dhillon95a}
Dhillon, V.~S. \& Marsh, T.~R. 1995, MNRAS, 275, 89

\bibitem[{Duquennoy \& Mayor(1991)}]{duquennoy91}
Duquennoy, A. \& Mayor, M. 1991, A\&A, 248, 485

\bibitem[{Harris {et~al.}(1987)Harris, Lambert, \& Goldman}]{harris87}
Harris, M.~J., Lambert, D.~L., \& Goldman, A. 1987, MNRAS, 224, 237

\bibitem[{Harris {et~al.}(1988)Harris, Lambert, \& Smith}]{harris88}
Harris, M.~J., Lambert, D.~L., \& Smith, V.~V. 1988, ApJ, 325, 768

\bibitem[{Iben \& Livio(1993)}]{iben93}
Iben, I. \& Livio, M. 1993, PASP, 105, 1373

\bibitem[{Livio(1996)}]{livio96}
Livio, M. 1996, in Evolutionary processes in Binary Stars, ed. R.~A. M.~J.
  Wijers, M.~B. Davies, \& C.~A. Tout (Dordrecht: Kluwer Academic Publishers),
  141

\bibitem[{O'Brien {et~al.}(2001)O'Brien, Bond, \& Sion}]{obrien01}
O'Brien, M.~S., Bond, H.~E., \& Sion, E.~M. 2001, ApJ, 563, 971

\bibitem[{Paczy\'{n}ski(1976)}]{paczynski76}
Paczy\'{n}ski, B. 1976, in Structure and Evolution of Close Binary Systems, ed.
  P.~P. Eggleton, S.~Mitton, \& W.~J.~A. J. (Dordrecht: Reidel), 75

\bibitem[{{Ramseyer} {et~al.}(1995){Ramseyer}, {Hatzes}, \&
  {Jablonski}}]{ramseyer95}
{Ramseyer}, T.~F., {Hatzes}, A.~P., \& {Jablonski}, F. 1995, AJ, 110, 1364

\bibitem[{Rappaport {et~al.}(1983)Rappaport, Joss, \& Verbunt}]{rappaport83}
Rappaport, S., Joss, P.~C., \& Verbunt, F.~A. 1983, ApJ, 275, 713

\bibitem[{Ritter \& Kolb(1998)}]{ritter98}
Ritter, H. \& Kolb, U. 1998, A\&AS, 129, 83

\bibitem[{Sarna {et~al.}(1995)Sarna, Dhillon, Marsh, \& Marks}]{sarna95}
Sarna, M.~J., Dhillon, V.~S., Marsh, T.~R., \& Marks, P.~B. 1995, MNRAS, 272,
  L41

\bibitem[{Spruit \& Ritter(1983)}]{spruit83}
Spruit, H.~C. \& Ritter, H. 1983, A\&A, 124, 267

\bibitem[{Wallace \& Hinkle(1996)}]{wallace96}
Wallace, L. \& Hinkle, K. 1996, ApJS, 107, 312

\bibitem[{{Webb} {et~al.}(2002){Webb}, {Naylor}, \& {Jeffries}}]{webb01}
{Webb}, N.~A., {Naylor}, T., \& {Jeffries}, R.~D. 2002, ApJ, 568, L45

\bibitem[{{Wright}(1994)}]{wright94}
{Wright}, G.~S. 1994, Experimental Astronomy, 3, 17

\bibitem[{{Young} \& {Nelson}(1972)}]{young72}
{Young}, A. \& {Nelson}, B. 1972, ApJ, 173, 653

\end{thebibliography}
\end{document}